\documentclass[useAMS,usenatbib]{mn2e}
\usepackage{graphicx}
\usepackage{natbib}
\usepackage{setspace}
\usepackage{deluxetable}
\newcommand{\kms}{~km\,s$^{-1}$}
\renewcommand{\deg}{$^{\circ}$}
\newcommand{\apm}{\,$\pm$\,}
\newcommand{\atimes}{\,$\times$\,}
\newcommand{\asim}{\,$\sim$\,}
\newcommand{\lt}{\,$<$\,}
\newcommand{\glong}{{\em{l}}}
\newcommand{\glat}{{\em{b}}}
\newcommand{\cz}{{\em{cz}}}
\newcommand{\msol}{\,$h^{-1}$~M$_{\sun}$}
\def\la{\mathrel{\hbox{\rlap{\hbox{\lower4pt\hbox{$\sim$}}}\hbox{$<$}}}}
\begin{document}

\title[Structures in the GA region]{Structures in the Great Attractor Region}

\author[Radburn-Smith et al.]{
\parbox{\textwidth}{D.~J.~Radburn-Smith$^1$, J.~R.~Lucey$^1$,
  P.~A.~Woudt$^2$, R.~C.~Kraan-Korteweg$^2$, F.~G.~Watson$^3$}\\
\vspace*{4pt}\\
$^1$Department of Physics, University of Durham, South Road, Durham
  DH1 3LE, UK\\
$^2$Department of Astronomy, University of Cape Town, Rondebosch 7700, South Africa\\
$^3$Anglo-Australian Observatory, Coonabarabran, NSW 2357, Australia}

\pagerange{\pageref{firstpage}--\pageref{lastpage}} \pubyear{2006}

\maketitle

\label{firstpage}

\begin{abstract}
To further our understanding of the Great Attractor (GA), we have
undertaken a redshift survey using the 2dF on the AAT. Clusters and
filaments in the GA region were targeted with 25 separate pointings
resulting in approximately 2600 new redshifts. Targets included poorly
studied X-ray clusters from the CIZA catalogue as well as the Cen-Crux
and PKS~1343-601 clusters, both of which lie close to the classic GA
centre. For nine clusters in the region, we report velocity
distributions as well as virial and projected mass estimates. The
virial mass of CIZA~J1324.7--5736, now identified as a separate
structure from the Cen-Crux cluster, is found to be
$\sim$3\atimes10$^{14}$~M$_{\sun}$, in good agreement with the X-ray
inferred mass. In the PKS~1343-601 field, five redshifts are measured
of which four are new. An analysis of redshifts from this survey, in
combination with those from the literature, reveals the dominant
structure in the GA region to be a large filament, which appears to
extend from Abell S0639 (\glong\,=\,281\deg, \glat\,=\,+11\deg) to
(\glong~$\sim$~5\deg, \glat~$\sim$~--50\deg), encompassing the
Cen-Crux, CIZA~J1324.7--5736, Norma and Pavo II clusters. Behind the
Norma Cluster at \cz\asim15000\kms, the masses of four rich clusters
are calculated. These clusters (Triangulum-Australis, Ara,
CIZA~J1514.6--4558 and CIZA~J1410.4--4246) may contribute to a
continued large-scale flow beyond the GA. The results of these
observations will be incorporated into a subsequent analysis of the GA
flow.
\end{abstract}


\section{Introduction}
Peculiar velocities are vital probes of the large scale mass
distribution in the local Universe that do not rely on the assumption
that light traces mass. Early work by \cite{lyn88} made the unexpected
discovery of a 600\kms~outflow towards Centaurus. This led to the idea
of a large, extended mass overdensity, nicknamed the Great Attractor
(GA), dominating the dynamics of the local Universe. Whilst many
studies have confirmed the presence of the GA \citep[e.g.][]{aar89},
the precise mass, position and extent of the overdensity remain
uncertain. \cite{lyn88} originally located the GA at (\glong, \glat,
\cz) $\sim$ (307\deg, +9\deg, 4350\apm350\kms) with a mass of
5.4\atimes10$^{16}$~M$_{\sun}$. However a subsequent study by
\cite*{kol95} placed the GA peak at (320\deg, 0\deg, 4000\kms), whilst
\cite{ton00} favoured an even closer locale at (289\deg, +22\deg,
3200\apm260\kms) and a mass approximately six times smaller
($\sim$8\atimes10$^{15}$~M$_{\sun}$). As the GA lies in the Zone of
Avoidance (ZoA), foreground extinction and high stellar contamination
have hampered studies of the underlying galaxy distribution. Recently,
however, several key results have emerged.

The Norma cluster (Abell 3627), located at (325\deg, --7\deg,
4848\kms), is now recognised to be comparable in mass, richness and
size to the Coma cluster \citep{kra96}. Lying $\sim$9\deg~from the
\cite{kol95} location of the GA, the cluster has been identified as a
likely candidate for the `core' of the overdensity
\citep{wou98}. Furthermore, it has been suggested that the GA is a
`Great Wall' like structure that extends from low galactic latitudes,
encompassing the Pavo II \citep[332\deg, --24\deg, 4200\kms,][]{luc88}
and Norma clusters before bending over and continuing towards
\glong\asim290\deg~\citep{kra94,wou97,wou04}. This connection has been
labelled the Norma supercluster \citep{fai98b} and constitutes the
major structure in the GA region (defined here as
280\deg\lt\glong\lt360\deg, --45\deg\lt\glat\lt+30\deg,
3000\,$<$\cz\lt7000\kms).

The richness of such connective structures in the region have been
highlighted by recent blind HI surveys in the southern sky
(Kraan-Korteweg et al. 2005b; Koribalski 2005; Henning et
al. 2005)\nocite{kra05b, kor05, hen05}. Because the ZoA is effectively
transparent to 21 cm radiation, these surveys are able to trace the
full extent of the local large-scale filaments as they pass through
the plane. Notably, between galactic latitudes of \mbox{--5\deg} and
+5\deg, \cite{hen05} find evidence for an extension of the Norma
supercluster at \cz\asim5000\kms, running from \glat\,=\,300\deg~to
340\deg.

The X-ray selected `Clusters In the Zone of Avoidance' (CIZA) project
has revealed several new X-ray clusters at low galactic latitudes
(\citealt*{ebe02}; \citealt{koc05}).  In the GA region, this survey has
identified CIZA~J1324.7--5736 as another potentially sizeable
contributor to the GA's mass. Lying at (307\deg, +5\deg, 5700\kms)
this cluster has been associated with the overdensity previously
identified as the Cen-Crux cluster \citep{wou98}. X-ray measurements
suggest that the structure is comparable in mass to the Norma cluster
\citep{mul05}.

Another important cluster in the GA region may exist around
PKS~1343-601, an extremely strong radio source lying in the ZoA
\citep{kra99}. The host galaxy is a large E0 \citep*{lau77,wes89}
located at $\sim$~(310\deg, +2\deg, 3900\kms). Despite the lack of an
associated X-ray source \citep{ebe02}, recent near-infrared surveys
are consistent with the presence of an intermediate mass cluster
centred on the radio source (Kraan-Korteweg et al. 2005a; Schr\"oder
et al. 2005; Nagayama et al. 2004)\nocite{kra05a,sch05,nag04}.

Attempts to analyse the extent and mass of the GA from peculiar
velocity measurements have remained inconclusive. To date, no clear
sign of any backside infall has been detected
\citep*{mat92,hud94}. This has been attributed to a continuing high
amplitude flow, possibly due to the gravitational pull of the Shapley
supercluster \citep[SSC, ][]{sca89,ray89,bra99,hud04}. Centred on Abell 3558
(312\deg, 31\deg, 14500\kms), the SSC is an extremely rich
concentration of galaxies. Dynamical analysis by \cite{rei00} of the
collapsing core of the SSC, indicates that the mass contained within
the central 8\,{\em{h}}$^{-1}$~Mpc is between 2\atimes10$^{15}$ and
1.3\atimes10$^{16}$\msol. However different estimates of the SSC's
mass, derived from various surveys of the region, vary significantly
due to differing assessments of the extent and geometry of the
structure \citep[see][]{bar00}. Furthermore, recent analysis suggests
that intercluster galaxies may compose up to two thirds of the SSC's
mass, thus severely biasing previous estimates based solely on summed
cluster masses \citep{pro05}. Accounting for all the galaxies in their
285 deg$^2$ survey of the SSC, \cite{pro05} estimate an enclosed mass
of 5\atimes10$^{16}$\msol.

This uncertainty in the relative masses of the GA and the SSC has led
to much dispute over the predicted source of the bulk flow observed in
the local Universe and hence the source of the Local Group's (LG) own
motion. \cite{ett97} and \cite{row00} estimated that the SSC was only
responsible for approximately 5 per cent of the LG's motion. However,
\cite{bar00} placed the contribution closer to $\sim$15 per cent
whilst others have advocated values of up to 50 per cent (e.g
\citealt{smi00}; \citealt*{luc05}; \citealt{koc05}).

In order to further understand the nature of the GA, and hence the
role it plays in the LG's motion, we have undertaken a redshift survey
with the Two-degree Field multi-fibre spectrograph (2dF). Targets
include five of the CIZA clusters (including the Cen-Crux cluster),
the \mbox{PKS 1343--601} region and over-densities located along the
proposed filamentary structures. We describe these observations and
present the redshift measurements in Section~\ref{sec:observations}
where we also discuss errors and completeness. Analysis of the
identified structures are presented in Section~\ref{sec:structures}
and in Section~\ref{sec:summary} we summarise our findings.

\begin{table*}
\caption{Summary of 2dF observations. The (\glong, \glat) coordinates
  for each targeted field are listed. These are not necessarily
  identical to the coordinates of cluster centres, as small
  adjustments were made to maximise the number of galaxies available
  to fibres in each field.}
\label{tab:targets}
\begin{tabular}{llrrrlr}
\hline
Field No. & Target & \multicolumn{1}{c}{\glong} & \multicolumn{1}{c}{\glat} & Exposure length (s) &
UT Date & No. Redshifts\\
\hline
1 & Cen-Crux/CIZA J1324.7--5736 -- 1 & 307.4 & 4.9 & 3\atimes900 & 2004 Feb 29 & 46\\
2 & Cen-Crux/CIZA J1324.7--5736 -- 2 & 305.4 & 5.1 & 3\atimes900 & 2004 Feb 29 & 51\\
3 & Cen-Crux/CIZA J1324.7--5736 -- 3 & 305.1 & 7.1 & 3\atimes900 & 2004 Feb 29 & 40\\
4 & Cen-Crux/CIZA J1324.7--5736 -- 4 & 304.6 & 9.4 & 3\atimes900 & 2005 Jun 9 & 87\\
5 & PKS 1343-601 & 309.7 & 2.3 & 7\atimes900 & 2004 Feb 29 & 5\\
6 & Abell S0639 & 281.3 & 10.7 & 3\atimes1200 & 2004 Feb 29 & 174\\
7 & Triangulum-Australis/CIZA J1638.2--6420 & 324.7 & --11.7 & 3\atimes900 & 2005 Jun 8 & 252\\
8 & Ara/CIZA J1653.0--5943 & 329.2 & --9.8 & 3\atimes900 & 2005 Jun 8 & 179\\
9 & Cluster 1 & 314.3 & 13.9 & 3\atimes900 & 2005 Jun 8 & 225\\
10 & CIZA J1514.6--4558 & 327.3 & 10.2 & 3\atimes1200 & 2005 Jun 7 & 226\\
11 & CIZA J1410.4--4246 & 317.9 & 17.8 & 3\atimes900 & 2005 Jun 8 & 182\\
12 & Filament 1 & 296.3 & 9.1 & 4\atimes900 \& 1\atimes712 & 2005 Jun 8 & 135\\
13 & Hydra-Antlia Extension 1 & 281.8 & --6.2 & 3\atimes900 & 2005 Jun 9 & 91\\
14 & Hydra-Antlia Extension 2 & 280.6 & --7.8 & 3\atimes900 & 2005 Jun 9 & 126\\
15 & Filament 2 & 300.4 & 9.0 & 3\atimes900 & 2005 Jun 9 & 83\\
16 & Filament 3 & 299.8 & 6.9 & 3\atimes900 & 2005 Jun 9 & 50\\
17 & Filament 4 & 312.5 & 5.0 & 4\atimes900 & 2005 Jun 8 & 60\\ 
18 & Filament 5 & 316.6 & 8.1 & 3\atimes900 & 2005 Jun 9 & 70\\
19 & Filament 6 & 312.9 & 9.0 & 3\atimes900 & 2005 Jun 9 & 101\\
20 & Filament 7 & 312.6 & 12.4 & 3\atimes900 & 2005 Jun 8 & 111\\ 
21 & Filament 8 & 351.0 & --22.6 & 3\atimes900 & 2005 Jun 8 & 146\\
22 & Filament 9 & 355.3 & --33.0 & 2\atimes900 & 2005 Jun 8 & 175\\
23 & Filament 10/RXC J1840.6-7709 & 317.7 & --25.5 & 3\atimes900 & 2005 Jun 9 & 156\\ 
24 & Filament 11/CIZA J1407.8--5100 & 315.0 & 10.2 & 3\atimes900 & 2005 Jun 9 & 91\\ 
25 & Cluster 2 & 322.3 & 13.6 & 3\atimes900 & 2005 Jun 9 & 155\\
26 & Ara/CIZA J1653.0--5943 -- repeat & 329.2 & --9.8 & 4\atimes900 & 2005 Jun 9 & 169\\
\hline
\end{tabular}
\end{table*}

\section{Observations and data reduction}
\label{sec:observations}

Observations were carried out in two runs on the 3.9m Anglo-Australian
Telescope (AAT). The 2dF was configured using the same set up as that
used for the 2dF Galaxy Redshift Survey \citep[2dFGRS][]{col01}. This
included using the 300B gratings with the 1024\atimes1024 24~$\mu$m
pixels on the Tektronix CCDs, resulting in a dispersion of
178.8~\AA~mm$^{-1}$ or 4.3~\AA~pixel$^{-1}$. At the centre of the
chip, the FWHM of the focus is about 2 pixels, hence the typical
spectral resolution is 9~\AA. Additionally, a central wavelength of
5800~\AA~was chosen to cover a range of about 3650--8050~\AA. Seeing
over the course of the two runs was $\sim$1--1.5 arcsec.

In total, we observed 25 separate fields as listed in
Table~\ref{tab:targets}. A repeat observation of one field was also
taken in order to assess systematics. Field centres were chosen to
maximise the number of targeted galaxies, whilst fully encompassing
known clusters and noticeable overdensities. Target galaxies were
taken from the Two Micron All Sky Survey Extended Source Catalogue
\citep[2MASS~XSC,][]{jar00} and the NASA Extragalactic Database
(NED). Additional targets in the Cen-Crux and PKS 1343--601 fields
were identified using {\em{J}}, {\em{H}} and {\em{Ks}} observations
taken with the 1.4~m InfraRed Survey Facility \citep[IRSF,
][]{nag04,nag05} and I-band images from the Wide Field imager (WFI) at
the ESO 2.2m telescope at La Silla (Kraan-Korteweg et
al. 2005a)\nocite{kra05a}. Suitable guide stars were selected from the
Tycho 2 catalogue \citep{hog00}. 2MASS positions were used for both
targets and guide stars, with counterparts identified from the 2MASS
Point Source Catalogue \citep{cut03} for sources with no equivalent
2MASS~XSC position.

After acquiring each target field, a flat field and an arc exposure,
using copper-argon and copper-helium lamps, were taken for fibre
identification and wavelength calibration. Three 900\,s exposures of
the fields yielded signal to noise ratios of $\sim$15--30. However,
seven 900\,s exposures of targets in the PKS~1343--601 field achieved
an average S/N ratio of only $\sim$~5 due to high galactic extinction
({\em{A}}$_B$\asim10).

The data was reduced using the {\textsc{2dfdr}} automatic data
reduction program as described in \cite{col01}. The default settings
were used with the exception of the use of sky flux methods for fibre
throughput calibration, as no off-sky measurements were taken. Once
reduced, redshifts were measured using the {\textsc{runz}} program
developed for the 2dFGRS \citep[also described in][]{col01}. This
program uses the \cite{ton79} technique to cross correlate nine
templates with the observed spectra in order to obtain the best
absorption redshift. Where available, the program also determines
emission redshifts by matching O~{\textsc{ii}}, H$\beta$,
O~{\textsc{iii}}, H$\alpha$, N~{\textsc{ii}} and S~{\textsc{ii}}
features.

\subsection{Redshifts}
\label{sec:redshifts}

A total of 3053 redshifts were measured, 2603 of which are not listed
in NED (as of 2006 February 15). Table~\ref{tab:table} lists a
representative sample of the complete table which can be found
online\footnote{The full table will be published online in the
electronic edition of this journal and on the CDS}.
\addtocounter{footnote}{1}\footnotetext{As of March 2006, seven
galaxies are contained in neither the NED or 2MASS~XSC catalogues: Two
galaxies identified with the prefix KKOWA were found from ESO 2.2m WFI
{\em{I}}-band observations around PKS~1343-601 (Kraan-Korteweg et
al. 2005a)\nocite{kra05a}, two galaxies, labelled NNSW, are taken from
NIR IRSF observations around Cen-Crux \citep{nag05} and a further
three galaxies, labelled DJRS, are new identifications from searches
of DSS images.}\addtocounter{footnote}{-2}

\begin{table}
\caption{A representative sample of the full table published
  online. Both heliocentric absorption and emission redshifts are
  listed where measured. Column 1 lists the galaxy
  identification. The 2MASS~XSC name is given first
  and then the equivalent NED
  identification\protect\footnotemark. J2000 equatorial coordinates
  are listed as either part of the name of the target or after the
  colon in the first column. The 2MASS {\em{J}}-band magnitude
  (\texttt{j\string_m\string_ext}), extrapolated from a fit to the
  radial surface brightness profile, is listed in column 2 where
  available. Columns 3 and 4 list the heliocentric velocities ($cz$
  \kms) identified through absorption and emission features
  respectively. As discussed below, the uncertainty on each
  measurement is $\pm$\,85\kms.}
\label{tab:table}
\begin{tabular}{lrrr}
\hline
Name & {\em{J}}$_{\rm{Ext}}$ & \cz$_{ab}$ & \cz$_{em}$ \\
\hline
\multicolumn{4}{c}{\bf{Field: 1 (RA:201.17\deg~Dec:-57.68\deg~\glong:307.78\deg~\glat:4.90\deg)}}\\
 2MASX J13184671-5804502 & 13.00 & & 14774 \\ 
 2MASX J13190643-5744311 & 12.38 &  5552 &  5507 \\ 
 2MASX J13200919-5725561 & 12.15 &  4578 & \\ 
 2MASX J13203723-5752421 & 11.57 &  5469 & \\ 
 2MASX J13211580-5827564 & 12.71 &  6155 & \\ 
 2MASX J13212199-5718084 & 14.11 &  6949 &  6835 \\ 
 2MASX J13220594-5728001 & 12.15 &  5706 & \\ 
 2MASX J13230235-5732041 & 12.15 &  5204 & \\ 
 2MASX J13230489-5740301 & 12.38 &  5841 &  5798 \\ 
 2MASX J13231390-5709190 & 12.28 &  5763 & \\ 
 2MASX J13232993-5744020 & 13.22 &  6068 & \\ 
NNSW71:J13233545-5747205 & & & 32701 \\ 
 2MASX J13233881-5807500 & 12.19 &  5444 & \\ 
 2MASX J13234325-5731460 & 13.20 &  6433 & \\ 
 2MASX J13234503-5742550 & 12.65 &  4426 & \\ 
 2MASX J13235263-5723200 & 12.29 &  5967 &  5870 \\ 
... &... & ...\\
\hline
\end{tabular}
\end{table}

Emission line redshifts are reported for approximately 32 per cent of
the sample, whilst absorption line based cross correlation redshifts are
available for $\sim$\,96 per cent. For the $\sim$\,27 per cent
identified through both absorption and emission features, the
absorption redshift is found to be larger on average by
$\sim$\,58\kms. This difference, which is usually attributed to gas
outflows, is consistent with offsets found in other galaxy surveys
\citep[e.g.][]{cap98}.

\begin{figure}
  \begin{center}
    \resizebox{90mm}{!}{\includegraphics{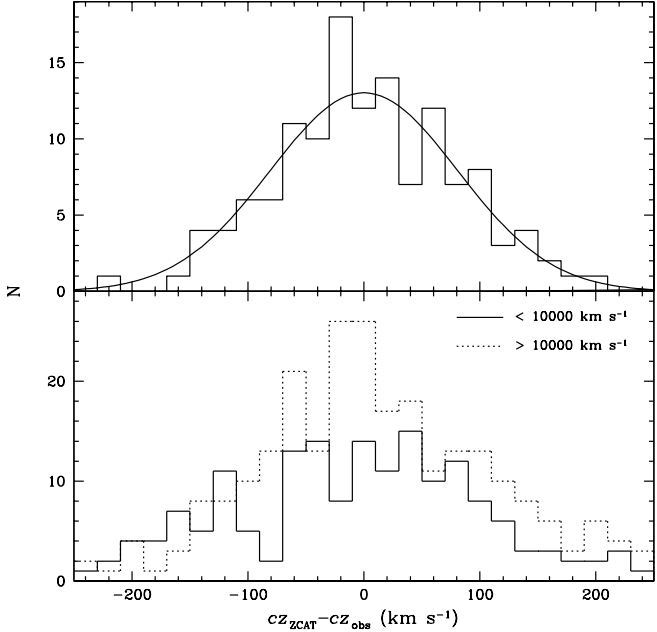}}
    \caption{The top panel shows the difference between repeat
      observations of the same field. A Gaussian fit to the dispersion
      yields a value of $\sigma$\,=\,114\kms, corresponding to a
      single measurement rms uncertainty of 81\kms. The bottom panel
      plots the difference between coincident measurements from the
      ZCAT catalogue. Histograms are plotted separately for data
      within 10000\kms~and for data beyond as coincident measurements
      primarily fall into two distinct velocity ranges around
      6000\kms~and 15000\kms. The mean offset of the points is
      +2\kms~and the scatter is consistent with an error of 89\kms~on
      our data points.}
    \label{fig:errors}
  \end{center}
\end{figure}

In order to assess the combined reliability of the observations and
data reduction, a repeat observation of one field
(Ara/CIZA~J1653.0--5943) was made. The difference between these
measurements (shown in the top panel of Fig.~\ref{fig:errors}) implies
an rms uncertainty on a single measurement of 81\kms.

The lower panel of Fig.~\ref{fig:errors} shows the residual
differences between our data and those from ZCAT \citep[][, 2005
November 27 edition]{huc92}. Coincident galaxies between the
catalogues were found through name matching and searching for
separations of less than 4 arcsec. For the resulting 433 galaxies, a
negligible offset of only +2\kms~is found. A value of
$\chi^2_{\nu}\sim$\,1 is achieved by adopting an uncertainty of
89\kms~on our values and using the quoted ZCAT errors, which in the
absence of multiple measurements are taken directly from the original
source. At \cz\asim6500\kms, the comparison exhibits an excess of
negative values (i.e. ZCAT values significantly lower than the
redshifts reported here). This can be attributed to the inclusion in
ZCAT of redshifts for galaxies in Abell S0639 as measured by
\cite{ste96}. These measurements are offset from the rest of the ZCAT
catalogue by~$\sim$\,--140\kms, causing the enhancement around this
value in the residual histogram that represents comparisons within
10000\kms.

Comparison of the 221 galaxies in common with the 6dF Galaxy Survey
\citep[6dfGS 2DR,][]{jon05} indicates an error of 94\kms~with a mean
offset of +3\kms. Whilst analysis of the 96 galaxies also observed by
\cite{wou04} yields an 89\kms~uncertainty and +19\kms~offset. Hence,
as with the 2DFGRS \citep{col01}, we adopt an underlying random error
of 85\kms~on all our measurements.

\begin{figure}
  \begin{center}
    \resizebox{90mm}{!}{\includegraphics{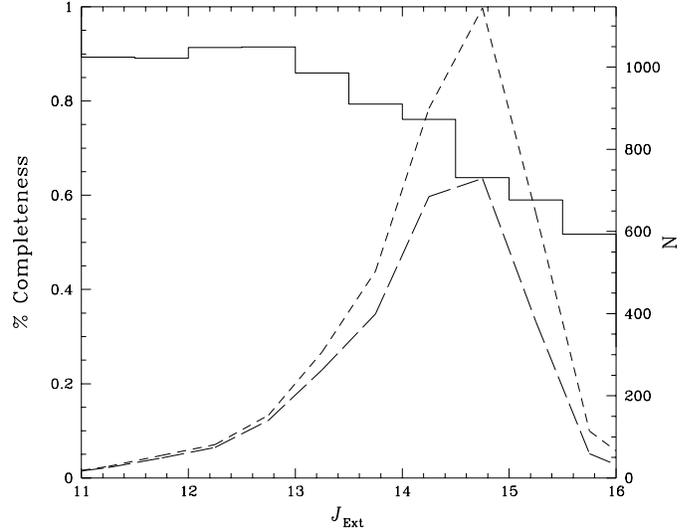}}
    \caption{The completeness of targeted galaxies. The solid
      histogram indicates the percentage of targeted galaxies in each
      0.5~mag bin for which a reliable redshift was discernible. The
      short and long dashed lines show respectively the total number
      of galaxies targeted and the number actually recorded in each
      corresponding bin.}
    \label{fig:complete}
  \end{center}
\end{figure}

The completeness of the observed 2MASS galaxies as a function of the
extrapolated {\em{J}}-band magnitude is shown in
Fig.~\ref{fig:complete}. The vast majority of targeted galaxies lie in
the range 12\lt{\em{J}}$_{\rm{Ext}}$\lt16~mag. Typically 10 per cent
of these yield no reliable redshift due to dominant stellar
contamination. Hence this survey has good completeness to
{\em{J}}\,=\,13~mag, after which a steady decline is observed down to
an effective completeness of $\sim$60 per cent for the faintest
galaxies at {\em{J}}\,$>$\,16~mag. To illustrate the depth of the
survey we calculate the characteristic magnitude at the distance of
the GA and the SSC. By fitting a Schechter function to the combined
2dFGRS/2MASS infrared catalogue, \cite{col01b} find a magnitude
corresponding to the characteristic luminosity {\em{L}}$^\star$ of
$M^\star_J-5\log h=-22.36\pm0.02$. Using this value we find an
apparent magnitude of {\em{J}}\asim11\,mag at the GA
(\cz\asim4500\kms) and $\sim$\,13.5\,mag at the SSC
(\cz\asim14500\kms). Around the Norma cluster and the SSC, extinction
is typically {\em{A}}$_J$\asim0.17 and 0.05\,mag respectively.

\begin{figure*}
  \begin{center}
    \resizebox{180mm}{!}{\includegraphics{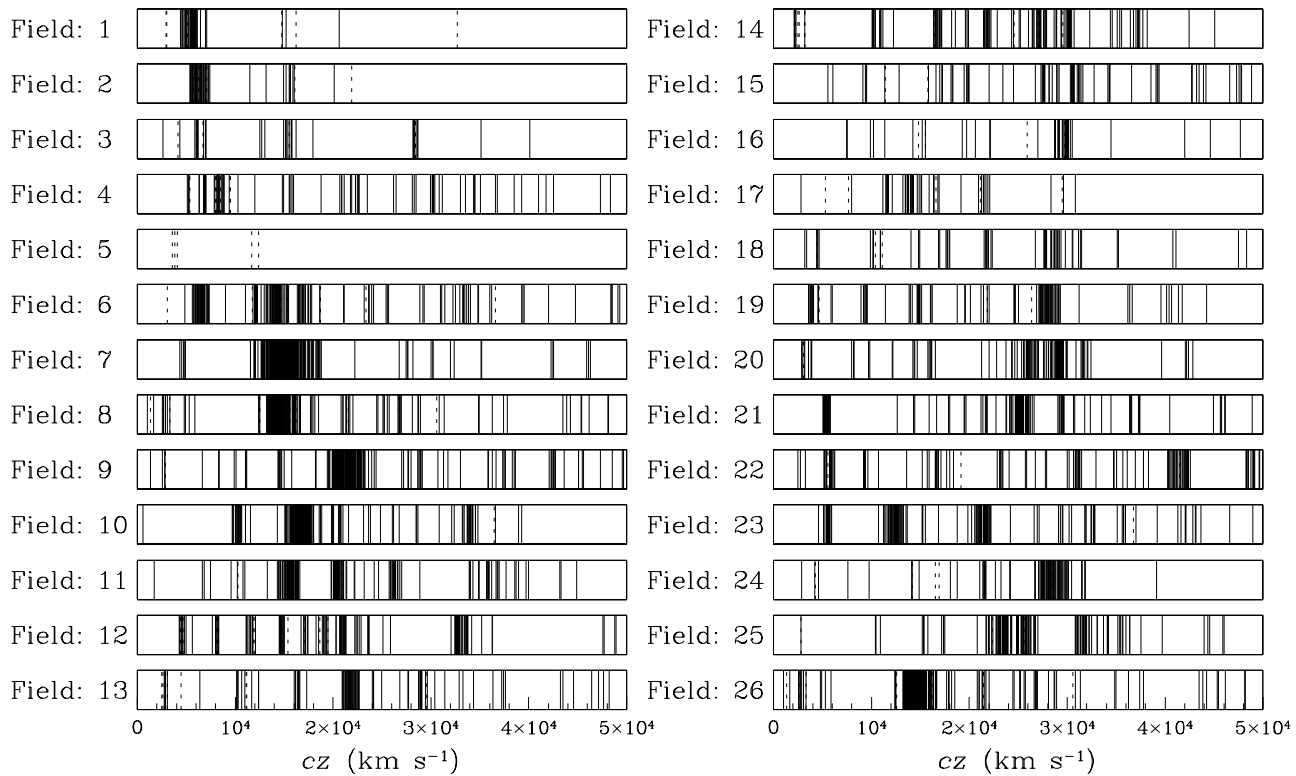}}
    \caption{Distribution of radial velocities in each of the 26
    targeted fields listed in Table~\ref{tab:targets}. Dashed lines refer to redshifts derived through
    observed emission lines, whilst solid lines indicate measurements
    made via cross-correlation with template spectra. Note that field
    26 is a repeat observation of field 8.}
    \label{fig:distributions}
  \end{center}
\end{figure*}

\section{Large-Scale Structures in the GA/SSC direction}
\label{sec:structures}

The redshift distribution for each of the surveyed fields is shown in
Fig.~\ref{fig:distributions}. Immediately obvious are the large
over-densities in fields 1,2 \& 6--11 corresponding to the targeted
clusters. The structures in which these clusters are embedded are also
apparent in many of the fields as features at redshifts of around
2000--6000\kms~and $\sim$15000\kms, corresponding to the GA and SSC
respectively.

\subsection{Review of large scale structures}
\label{sec:overview}

\begin{figure*}
  \begin{center}
    \resizebox{145mm}{!}{\includegraphics{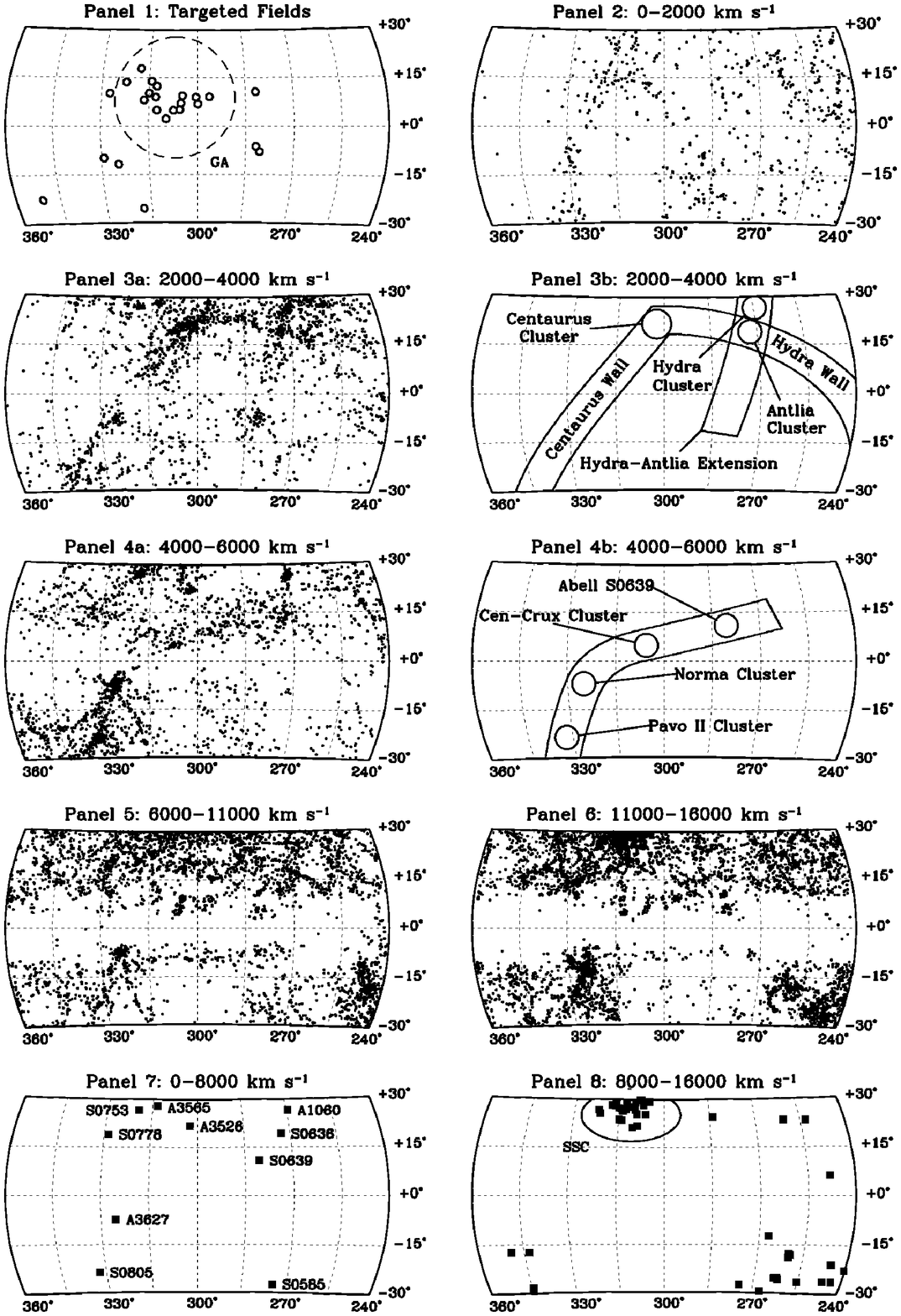}}
    \caption{Aitoff projections of redshift slices containing galaxies
      in the range 240$^{\circ}$\lt\glong\lt360\deg~and
      --30$^{\circ}$\lt\glat\lt+30\deg~from this survey and the NED
      database (as of 2006 February 15). The projected circles in
      the first panel represent the actual size of each 2dF target
      field located in the region. The dashed circle represents the
      core radius used in the spherical GA model of Faber \& Burstein
      (1988) centred on (306\deg, +9\deg). Panels 3b and 4b illustrate
      the key features observed in the corresponding redshift
      slices. Abell clusters within 8000\kms~are labelled in Panel 7,
      whilst in panel 8, Abell clusters between 8000 and 16000\kms~are
      plotted and the clusters composing the SSC are indicated.}
    \label{fig:sky}
  \end{center}
\end{figure*}

\nocite{fab88} The number of redshifts known in the GA and SSC region
have greatly increased with the recent completion of surveys such as
FLASH \citep{kal03}, 6dFGS, the SSC study of \cite{pro05} and the
`extragalactic large-scale structures behind the southern Milky Way'
project (\citealt*{kra94}; \citealt*{fai98b}; \citealt*{wou99};
\citealt{wou04}). Together with our measurements, we use these recent
surveys to assess the large-scale structures traced by the galaxies in
this important region. Fig.~\ref{fig:sky} plots the combined projected
distribution of the redshifts. The first panel identifies the 2dF
fields observed by this survey. The majority of fields lie in regions
outside the 6dFGS survey limit (i.e. \glat$<|10|$\deg) and
predominantly near 2MASS over-densities close to the classic GA
centre. Abell clusters are identified in the last two panels. whilst
the remaining panels present the data in successive redshift slices,
which contain the following relevant structures:

{\bf{\cz~$\le$~2000\kms:}} In this panel, a line of galaxies crossing
the Galactic plane at \glong\,=\,280\deg~and extending to the centre
of the Virgo Cluster (off the panel at \glong\,=\,280\deg,
\glat\,=\,+74\deg) is clearly seen. These belong to the Virgo
Supercluster, which encircles the entire sky and defines the
Supergalactic Plane. The smaller Fornax Wall is also seen here face-on
\citep{fai98a}. It appears as a filament of galaxies running from the
Fornax cluster (237\deg, --54\deg) and crossing the Galactic plane at
\glong\,=\,295\deg. The extension of these filaments through the ZoA
is traced by the HI galaxies from surveys based on the HI Parkes
All-Sky Survey \citep[HIPASS, ][]{bar01}, most notably the HIPASS
Bright Galaxy Catalogue \citep{kor04} and the deep HIPASS catalogue
\citep[HICAT, ][]{mey04}.

{\bf{2000~$<$~\cz~$\le$~4000\kms:}} Immediately apparent in the third
panel, is the Centaurus cluster (Abell 3526) lying at (302\deg,
+22\deg). Extending down from this cluster and through the galactic
plane is the Centaurus Wall. This wall crosses a large part of the
southern sky and is one of the most prominent features in all-sky maps
of galaxies within 6000\kms~\citep{fai98a}. As we lie close to the
plane of the Centaurus Wall, the structure is seen edge-on
\citep{fai98a}.

Almost perpendicular to the Centaurus Wall is the Hydra Wall
\citep{fai98a}. This is seen here as a filament of galaxies reaching
out from the Centaurus cluster, through the Hydra (270\deg, +27\deg)
and Antlia (273\deg, +19\deg) clusters before heading on to the Puppis
cluster \citep[240\deg, 0\deg, ][]{lah93} and down towards (210\deg,
--30\deg).

The Hydra-Antlia extension \citep{kra94} forms a third filamentary
structure in this slice. From the Hydra cluster, this feature passes
through the Antlia cluster, crosses the Galactic plane at
\glat~=278\deg~and ends in a group of galaxies at (280\deg,
--8\deg). \cite{kra94} suggested that an overdensity of galaxies,
named the Vela overdensity and located at (280\deg, +6\deg), formed
part of the Hydra-Antlia extension. However subsequent observations of
this group have revealed that it lies significantly behind the
extension at \cz\,=\,6000\kms~\citep{kra95}.

\begin{table*}
\caption{Parameters for the fits to the velocity distributions of the
  observed clusters as detailed in Section~\ref{sec:dispersions}.}
\label{tab:parameters}
\begin{tabular}{lr@{\apm}lr@{\apm}lrrlrr}
\hline
Cluster Name & \multicolumn{2}{c}{$\bar{v}$} & \multicolumn{2}{c}{$\sigma$} &
\multicolumn{1}{c}{M$_{\rm{Virial}}$} & \multicolumn{1}{c}{M$_{\rm{Projected}}$} & \multicolumn{1}{c}{W} & \multicolumn{1}{c}{N} & \multicolumn{1}{c}{p}\\
 & \multicolumn{2}{c}{\kms} & \multicolumn{2}{c}{\kms} & \multicolumn{1}{c}{{\em{h}}$^{-1}$~M$_{\odot}$} & \multicolumn{1}{c}{{\em{h}}$^{-1}$~M$_{\odot}$} & & & \\
\hline
CIZA~J1324.7--5736 & 5570&92 & 618&72 & (3.5\apm1.0)\atimes10$^{14}$ & (3.9\apm0.7)\atimes10$^{14}$ & 0.9555 & 40 & 0.1176\\
Abell S0639A & 6501&61 & 405&40 &
(1.2\apm0.3)\atimes10$^{14}$ & (1.7\apm0.4)\atimes10$^{14}$ &
0.983 & 40 & 0.7987\\
Abell S0639B & 14125&66 & 412&39 &
(3.6\apm0.8)\atimes10$^{14}$ & (5.3\apm0.6)\atimes10$^{14}$ &
0.951 & 41 & 0.0648\\
Triangulum Australis & 15060&97 & 1408&67 &
(5.7\apm0.6)\atimes10$^{15}$ & (6.9\apm0.5)\atimes10$^{15}$ &
0.9855 & 220 & 0.0242\\
\hspace{6pt}(corrected) & 14898&90 & 1246&59 &
(4.4\apm0.4)\atimes10$^{15}$ & (5.4\apm0.4)\atimes10$^{15}$ &
0.9919 & 210 & 0.2945\\
Ara & 14634&76 & 881&48 & (2.0\apm0.3)\atimes10$^{15}$ & (2.6\apm0.2)\atimes10$^{15}$ & 0.9840 & 147 & 0.0850\\
CIZA J1514.6--4558 & 16715&50 & 601&35 & (1.2\apm0.1)\atimes10$^{15}$ & (1.5\apm0.1)\atimes10$^{15}$ & 0.9953 & 149 & 0.9145\\
CIZA J1410.4--4246A & 15574&63 & 497&40 & (5.2\apm0.9)\atimes10$^{14}$ & (6.2\apm0.8)\atimes10$^{14}$ & 0.9761 & 66 & 0.2328\\
CIZA J1410.4--4246B & 20463&53 & 345&37 & (5.3\apm1.3)\atimes10$^{14}$ & (7.5\apm0.8)\atimes10$^{14}$ & 0.9569 & 45 & 0.0922\\
Cluster 1 (Field 9) & 21445&78 & 925&52 & (3.1\apm0.3)\atimes10$^{15}$ & (3.8\apm0.3)\atimes10$^{15}$ & 0.9851 & 151 & 0.1023\\
Cluster 2 (Field 25) & \multicolumn{2}{c}{-} &
\multicolumn{2}{c}{-} & \multicolumn{1}{c}{-} & \multicolumn{1}{c}{-}
& 0.9685 & 85 & 0.0354\\
\hline
\end{tabular}
\end{table*}

{\bf{4000~$<$~\cz~$\le$~6000\kms:}} The fourth panel reveals the
massive Norma cluster of galaxies lying at (325\deg, --7\deg). Below
this and connected by a trail of galaxies is the Pavo II cluster
(Abell S0805, \glong\,=\,332\deg, \glat\,=\,--24\deg). Additionally,
two smaller filaments of galaxies are seen extending down from the
Norma cluster to both lower and higher galactic longitudes.

A less pronounced linear feature is also observed in this
panel. Continuing from the connection between the Pavo II and Norma
clusters, the structure extends across the Galactic plane and on
through CIZA J1324.7--5736 (307\deg, +5\deg) and the Cen-Crux
(305\deg, +5\deg) cluster before ending at Abell S0639 (281\deg,
+11\deg). Collectively, this structure is known as the `Norma
supercluster' \citep{wou97} and is discussed further in
Section~\ref{sec:normasupercluster}.

{\bf{6000~$<$~\cz~$\le$~11000\kms:}} The Norma cluster `finger of God'
is still evident in this panel. The linear feature at
\glat\,=\,--10\deg~that extends from this overdensity towards lower
galactic latitudes, is an artificial enhancement due to the survey
limit (\glat~$\la$~--10\deg) of the combined southern Milky Way survey
\citep{kra95,fai98b,wou99}. The Vela overdensity and continuation of
the Cen-Crux structure are both seen as distinct groups at (305\deg,
+6\deg) and (280\deg, +6\deg) respectively. Also present is the
Ophiuchus cluster \citep{has00,wak05} lying at the edge of the panel
(360\deg, +9\deg, 8500\kms).

{\bf{11000~$<$~\cz~$\le$~16000\kms:}} In the last panel, the massive
concentration of clusters that constitute the SSC becomes
apparent around (314\deg, +30\deg). Also visible are the large Ara
(329\deg, --10\deg) and Triangulum-Australis (325\deg, --12\deg)
clusters (lying almost directly behind the Norma cluster),
CIZA~J1514.6--4558 at (327\deg, +10\deg) and \mbox{CIZA~J1410.4--4246}
at (318\deg, +18\deg).

\subsection{Clusters}
\label{sec:dispersions}

Of great importance in studying the GA flow is an assessment of the
relative masses of the rich clusters in the region. Notably, the CIZA
survey has identified several new X-ray clusters in the GA direction.
We targeted six of these sources, which together with noticeable
overdensities in the 2MASS XSC, made up nine fields containing
possible clusters.

To determine if these systems were indicative of relaxed clusters,
their velocity dispersions, culled by an iterative 3-$\sigma$ clipping
procedure about their median, were tested for gaussianity. With no
prior on the mean or standard deviation, the Shapiro-Wilk W-statistic
\citep{sha65} is able to test the null hypothesis that data is indeed
sampled from a normal distribution. We accept this hypothesis if the
associated p-value, calculated via the analytical approach of
\cite{roy95}, is greater than 0.05.

If the W-statistic for a sample indicates that the redshifts were
taken from a normal distribution, the corresponding velocity
dispersion was determined using a method that includes measurement
errors on individual redshifts \citep*{dan80}. Uncertainties on the
derived values were calculated by bootstrap resampling.

The masses of the corresponding systems were calculated using the
classical virial mass estimator, defined by \cite{hei85} as
\begin{displaymath}
M_{\rm{vir}}=\frac{3\pi{N}}{2G}\frac{\sum_{i}(v_i-\bar{v})^2}{\sum_{i,j<i}R_{ij}^{-1}}\\
\end{displaymath}
where
\begin{displaymath}
R_{ij}=|R_i-R_j|
\end{displaymath}
is the projected galaxy separation. This virial method has been shown
to be a reliable first order approximation to the mass of a
dynamically relaxed system which is fully contained within the
observed field \citep[e.g. see][]{rin03}. The projected mass estimator
for each cluster was also calculated:
\begin{displaymath}
M_{\rm{proj}}=\frac{32}{N{\pi}G}\sum_{i}R_i(v_i-\bar{v})^2.
\end{displaymath}
Errors on both mass estimates were again assigned by bootstrap
resampling.  With their sample of nine clusters in the CAIRNS project,
\cite{rin03} find that the projected mass is only 1.18\apm0.05 times
greater than the estimated virial mass. Hence, given the expected
errors on the dispersions, the two estimators should be consistent.

\begin{figure*}
  \begin{center}
    \resizebox{180mm}{!}{\includegraphics{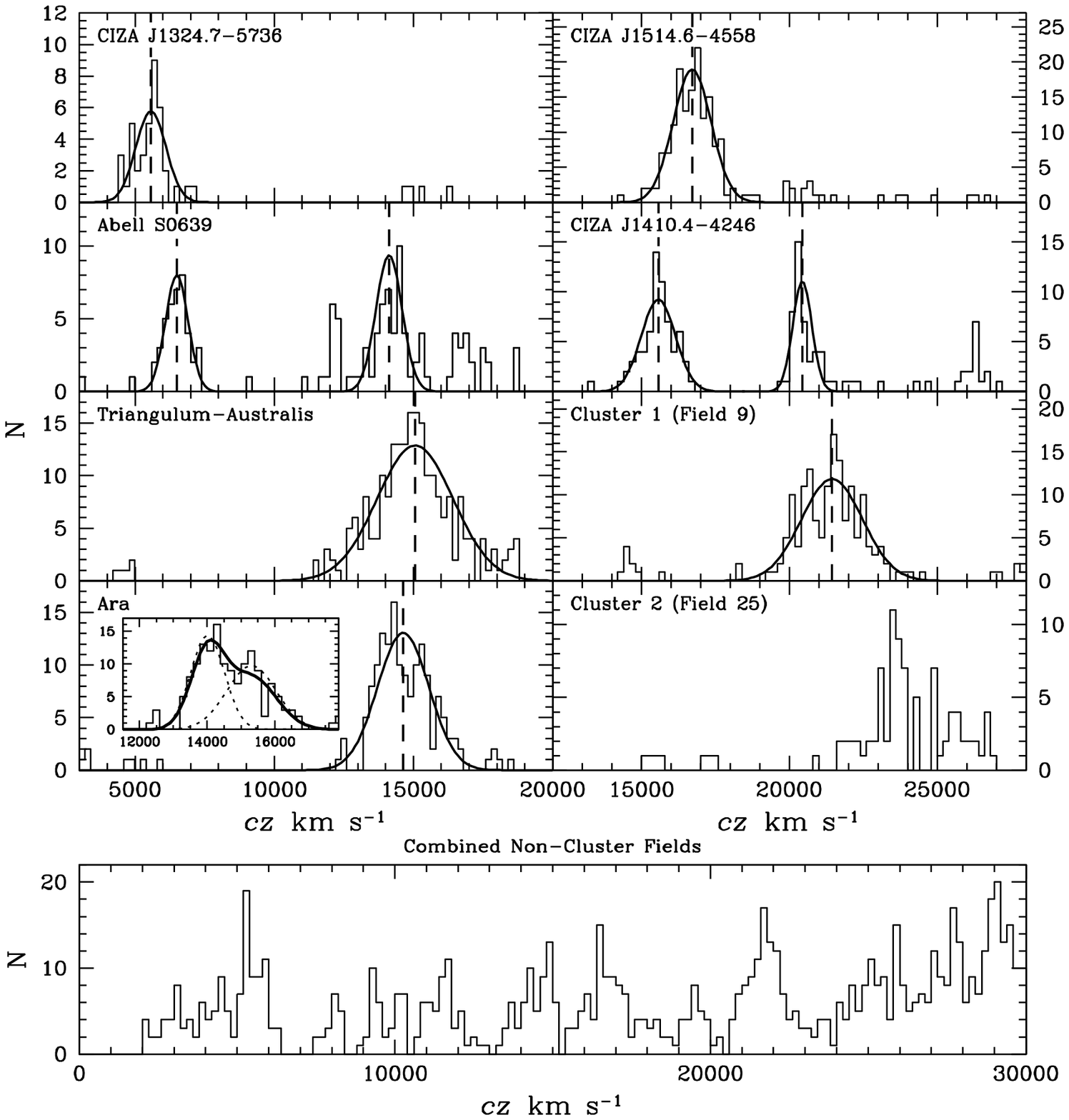}}
    \caption{The radial velocity dispersions and corresponding virial
    fits for the observed clusters are shown in the upper panel. The
    lower panel shows the combined velocity distribution for the
    11 non-cluster fields.}
    \label{fig:histograms}
  \end{center}
\end{figure*}

Table~\ref{tab:parameters} lists the mean redshift, velocity
dispersion, mass estimate, W-statistic and associated p-value for the
best fit to each of the observed clusters. These fits are plotted with
the corresponding velocity histograms in Fig.~\ref{fig:histograms}.

\subsubsection{Cen-Crux/CIZA J1324.7--5736}

\begin{figure*}
  \begin{center}
    \resizebox{150mm}{!}{\includegraphics{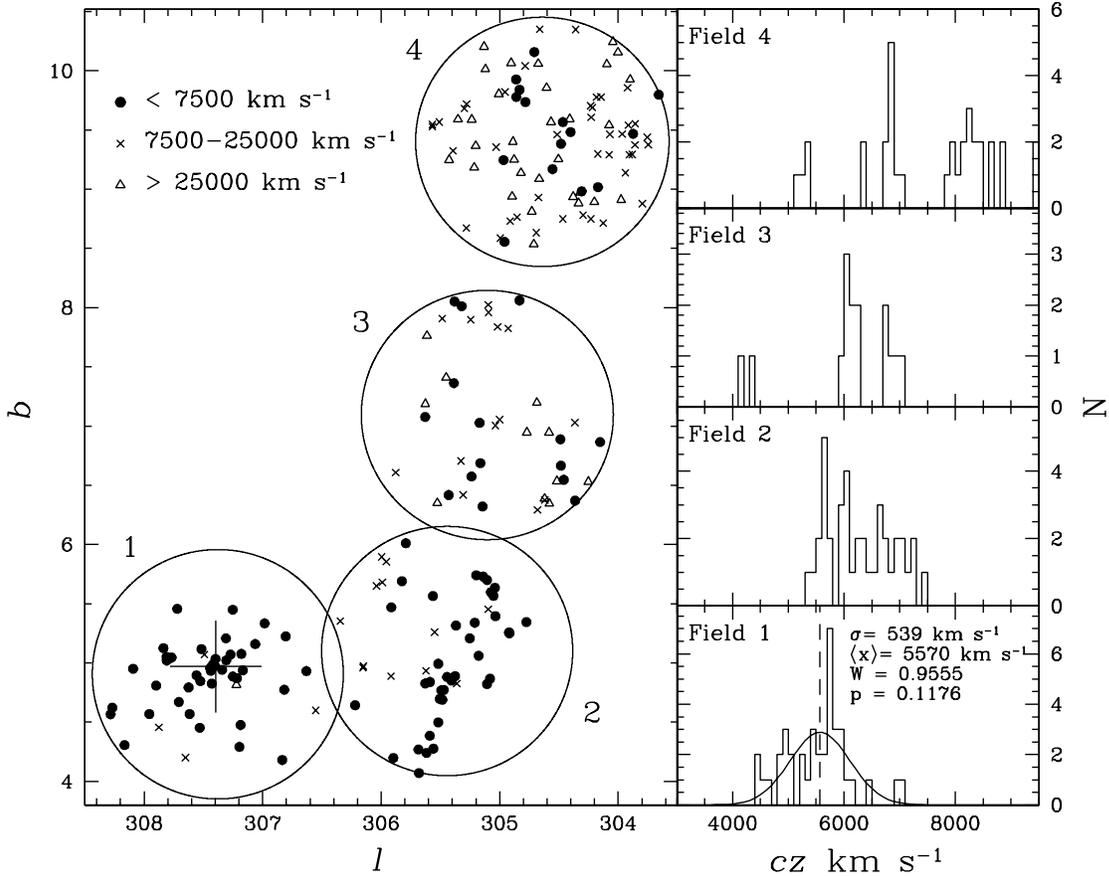}}
    \caption{Galactic longitude and latitude of galaxies measured by
      this survey for the four denoted 2dF fields in the Cen-Crux
      region. The large cross marks the centre of the X-ray source
      CIZA J1324.7--5736. The right hand panels show the corresponding
      velocity histograms for each of the fields between 3000 and
      9500\kms.}
    \label{fig:cencrux}
  \end{center}
\end{figure*}

Multi-object spectroscopy of the GA region revealed an overdensity of
galaxies at (305\deg, +5\deg, 6214\kms), which was named the Cen-Crux
cluster \citep{wou98, fai98b, wou04}. Later, an associated X-ray
cluster signature was detected by the CIZA survey at (307\deg,
+5\deg).  Preliminary analysis of the X-ray source
(CIZA~J1324.7--5736) suggested that it was comparable in mass to the
Norma cluster \citep{ebe02}.

We have observed one field centred on the X-ray source and three
further fields targeting the surrounding overdensities (see
Fig.~\ref{fig:cencrux}). Of the 223 identified redshifts in the
targeted fields, 110 are within 7500\kms. Two distinct structures are
observed within these fields.

\cite{ebe02} noted that the appearance of the X-ray emissions in the
region and their association with the brightest cluster galaxy WKK2189
(\cz\,=\,5585\kms), were suggestive of a dynamically relaxed
cluster. 40 of the observed galaxies are found to be associated with
the X-ray source. Shown in the Field 1 histogram on the right hand
side of Fig.~\ref{fig:cencrux}, the velocity dispersion of these
galaxies is 539\apm80\kms~centred on 5570\apm92\kms. The Shapiro-Wilk
test on this distribution yields a p-value of 0.1176 and the estimated
virial mass is (3.5\apm1.0)\atimes10$^{14}$\msol. Hence the
interpretation of a large relaxed cluster is supported here by the
observed Gaussian velocity distribution.

Comparison with the Norma cluster velocity dispersion of
897\kms~\citep{kra96} suggests that CIZA~J1324.7--5736 is
approximately 0.3--0.5 times as massive.  This is in agreement with
the \cite{mul05} comparison of XMM-Newton observations of
CIZA~J1324.7--5736 with the X-ray temperature of the Norma cluster
inferred by \cite{tam98}. Using the mass-temperature scaling
relations, they conclude that CIZA~J1324.7--5736 contains about a
third of the mass of the Norma cluster. A future study of the
extinction-corrected {\em{K$_S$}}-band luminosity function should
provide further constraints on the relative mass \citep{nag05}.

The second distinct feature observed in the fields is that of the
Cen-Crux overdensity itself. This appears as a filament like trail of
galaxies separated from the X-ray source both spatially on the sky and
in redshift. Although no connective structure is evident between this
overdensity and CIZA J1324.7--5736, their close proximity suggest that
they are gravitationally bound. As the structure is not dynamically
relaxed, virial theorem does not apply. However the extent of the
Cen-Crux structure and the number of galaxies contained within it
implies a mass similar to that of the CIZA~J1324.7--5736 cluster.

\subsubsection{PKS 1343--601}

PKS 1343--601 is the second brightest extragalactic radio source in
the southern sky \citep{mil52}. The associated galaxy, lying at
\citep[309.7\deg, +1.7\deg, 3872\kms,][]{wes89}, is a large elliptical
galaxy \citep{lau77,wes89}, typical of those found in cluster
cores. Hence it has been suggested that PKS~1343--601 may mark the
centre of another highly obscured ({\em{A$_B$}}\asim12) cluster
\citep{wou98,kra99}.

X-ray studies have yet to reveal any indication that such a hidden
cluster exists. No corresponding source is seen in the CIZA survey and
the point-like X-ray emissions reported by \cite{tas98} are consistent
with the radio lobes of PKS~1343--601 rather than intracluster gas
\citep{ebe02}. However in HIPASS observations, a small overdensity
around the radio galaxy has been detected (Kraan-Korteweg et
al. 2005b)\nocite{kra05b}. The nature of this overdensity has recently
been examined by three near-infrared surveys (Schr\"oder et al. 2005;
Kraan-Korteweg et al. 2005a; Nagayama et al. 2004)\nocite{sch05,
kra05a, nag04}. Through radial velocity studies, simulated
sky-projections and extrapolation of luminosity functions, these
surveys are all consistent with the notion of a low mass group or poor
cluster centred on PKS~1343--601.

Unfortunately, of the 84 targets we identified in the 2dF field, our
6300~s observation yielded only five reliable redshifts. Of these is a
reconfirmation of the redshift of PKS 1343--601. At 4065\apm85\kms,
this is in agreement with the \cite{wes89} value.  Of the other four
new measurements, all identified through emission lines, two are
located within 500\kms~of the radio galaxy. NWN2004 45 and NWN2004 51
are both taken from the \cite{nag04} catalogue and lie at 3861 and
3571\kms~respectively. These galaxies, together with those identified
both optically and in HI by \cite{sch05}, brings the number of
galaxies with known redshifts that are associated with the
PKS~1343--601 group up to 20.

\subsubsection{Abell S0639}

The Abell S0639 cluster, which lies at (281\deg, +11\deg), was first
studied in detail by \cite{ste94,ste97}, who for 32 galaxies measured
a mean velocity of 6194\apm78\kms~and a velocity dispersion of
431\apm52\kms. Using a sample of 40 galaxies with a mean
\cz\,=\,6501\apm61\kms, we find a similar dispersion of
409\apm55\kms. An additional feature is located in the same field,
offset from Abell~S0639 by 1.5\deg. At 14065\apm69\kms, the structure
lies at the same distance as the SSC and is not inconsistent with a
normal distribution (p-value\,=\,0.0648). The measured virial velocity
dispersion is 597\apm91\kms, corresponding to a mass of
(4.9\apm1.2)\atimes10$^{14}$\msol.

\subsubsection{Triangulum Australis, Ara, CIZA~J1638.2--6420, CIZA~J1514.6--4558 \& CIZA~J1410.4--4246}

In the extended CIZA catalogue, \cite{koc05} have identified several
X-ray sources located at {\em{z}}~$\sim$~0.05, which they suggest form
an extension to the SSC. In \cite{ebe02}, the same authors argue that
these clusters may be responsible for the observed continued flow
towards a point behind the GA. Of these sources we have targeted the
four largest: CIZA~J1638.2--6420 (the Triangulum-Australis cluster) at
(324.5\deg, --11.6\deg, 15060\kms), \mbox{CIZA~J1653.0--5943}
\citep[the Ara cluster, ][]{wou98} at (329.3\deg, \mbox{--9.9\deg},
14634\kms), CIZA~J1410.4--4246 (318.0\deg, 17.8\deg, 15574\kms) and
CIZA~J1514.6--5736 (327.3\deg, 10.0\deg, 16715\kms). All four
structures have clearly identified Gaussian velocity distributions
from which we are able to infer virial and projected masses as listed in
table~\ref{tab:parameters}. The Triangulum-Australis cluster yields a
noticeably low p-value (0.0242). This is due to the overdensity seen
in the right hand tail of the dispersion. Removing the 10 galaxies
with {\em{cz}}~$>$~18000\kms~from the field results in a more
respectable p-value of 0.2945 (listed as corrected in
table~\ref{tab:parameters}). With a corresponding virial mass of
(5.7\apm0.6)\atimes10$^{15}$\msol, this large cluster is similar in
mass to the Norma cluster.

Despite a p-value of 0.0850, the Ara cluster appears to display a
bimodal velocity distribution. Fitting two Gaussian profiles to the
data results in velocity dispersions of 498\apm68\kms~and
731\apm112\kms~centred on 14016\apm84\kms~and
15310\apm124\kms~respectively. These fits are shown in the inset to
the Ara cluster panel of Fig.~\ref{fig:histograms}. There is no
discernible separation in the projected sky distribution of the two
populations, hence they may be two infalling clumps collapsing along
the line of sight. A 7.5~ks {\em{ROSAT}} HRI observation of the
cluster supports this argument, as two distinct peaks, separated by
only 4~arcmin, were observed in the elongated X-ray emissions
\citep{ebe02}. Summed in quadrature, the two velocity dispersions are
similar to the dispersion of the overall fit (881\apm48\kms); hence,
even though virial theorem is not strictly applicable to such a
system, the mass derived from the total fit provides a likely upper
limit to the combined mass of the two clumps.

The results of the Shapiro-Wilk test for CIZA~J1514.6--4558 and
CIZA~J1410.4--4246 indicate that they are consistent with being
dynamically relaxed clusters as shown in
Fig.~\ref{fig:histograms}. Behind CIZA~J1410.4--4246 there appears a
second group with a velocity dispersion consistent with a normal
distribution. However with a skewness of 0.094, the mean distance and
the velocity dispersion of the feature are likely overestimated.

The Triangulum-Australis and Ara clusters are physically separated by
only $\sim$13.7~{\em{h}}$^{-1}$~Mpc and lie in approximately the same
plane as the CIZA~J1514.6--4558 and CIZA~J1410.4--4246
clusters. Abell~3558, the core of the SSC, lies only 38~Mpc from
CIZA~J1410.4--4246 and so these clusters may well form an extension to
the SSC. Nevertheless the presence of such large masses in close
proximity to each other has a sizeable influence on the X-ray based
dipole \citep{koc04}. The effects of this will be studied in more
detail by a subsequent paper.

\subsubsection{Additional Clusters}

Examination of 2MASS maps of the GA/SSC region reveals two further
overdensities centred on (314.5\deg, +13.7\deg) and (321.7\deg,
+13.4\deg). These were targeted in fields 9 (314.3\deg, +13.9\deg) and
25 (322.3\deg, +13.6\deg) respectively. Recently, \cite{koc05} have
reported the presence of an X-ray source, identified as
CIZA~J1358.7--4750, at (314.5\deg, +13.5\deg), coincident with the
structure in field~9. At \cz\,=\,21445\apm78\kms~this cluster is far
enough removed to have little influence ({\em{V}}$_{\rm{LG}}$\lt3\kms)
on local dynamics despite the large predicted mass
($\sim$~3\atimes10$^{15}$\msol).

As evident in the lower right panel of Fig.~\ref{fig:histograms}, The
galaxies between 21000 and 27000\kms~in field 25 are concentrated into
numerous sub-clumps loosely associated in a broad distribution. The
associated p-value of 0.0345 confirms that this is not consistent with
a dynamically relaxed cluster and hence we do not assign it a mass.

\subsection{The Extended Norma Supercluster}
\label{sec:normasupercluster}

\begin{figure}
  \begin{center}
    \resizebox{90mm}{!}{\includegraphics{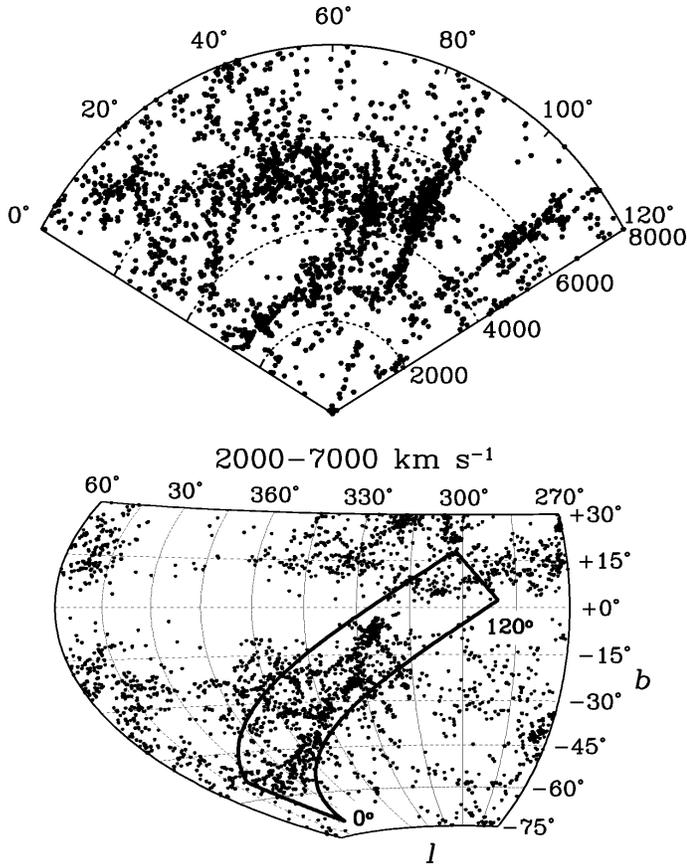}}
    \caption{The pieplot represents the radial distribution of
    galaxies along the projected rectangular strip shown in the lower
    panel. The strip covers a region 120\deg \atimes 10\deg,
    orientated to lie along the filament. From the Norma cluster,
    lying 86\deg~along the strip, the Norma supercluster is clearly
    seen as a wall of galaxies extending through the Pavo II cluster
    (at 71\deg) towards a point $\sim$\,20\deg~along the strip. The
    Centaurus Wall appears as a smaller connection of galaxies,
    running almost parallel to the Norma supercluster at 2600\kms. The
    void lying between the Norma supercluster and the Centaurus Wall
    is an extension of the massive Microscopium Void.}
    \label{fig:slice1}
  \end{center}
\end{figure}

\begin{figure}
  \begin{center}
    \resizebox{90mm}{!}{\includegraphics{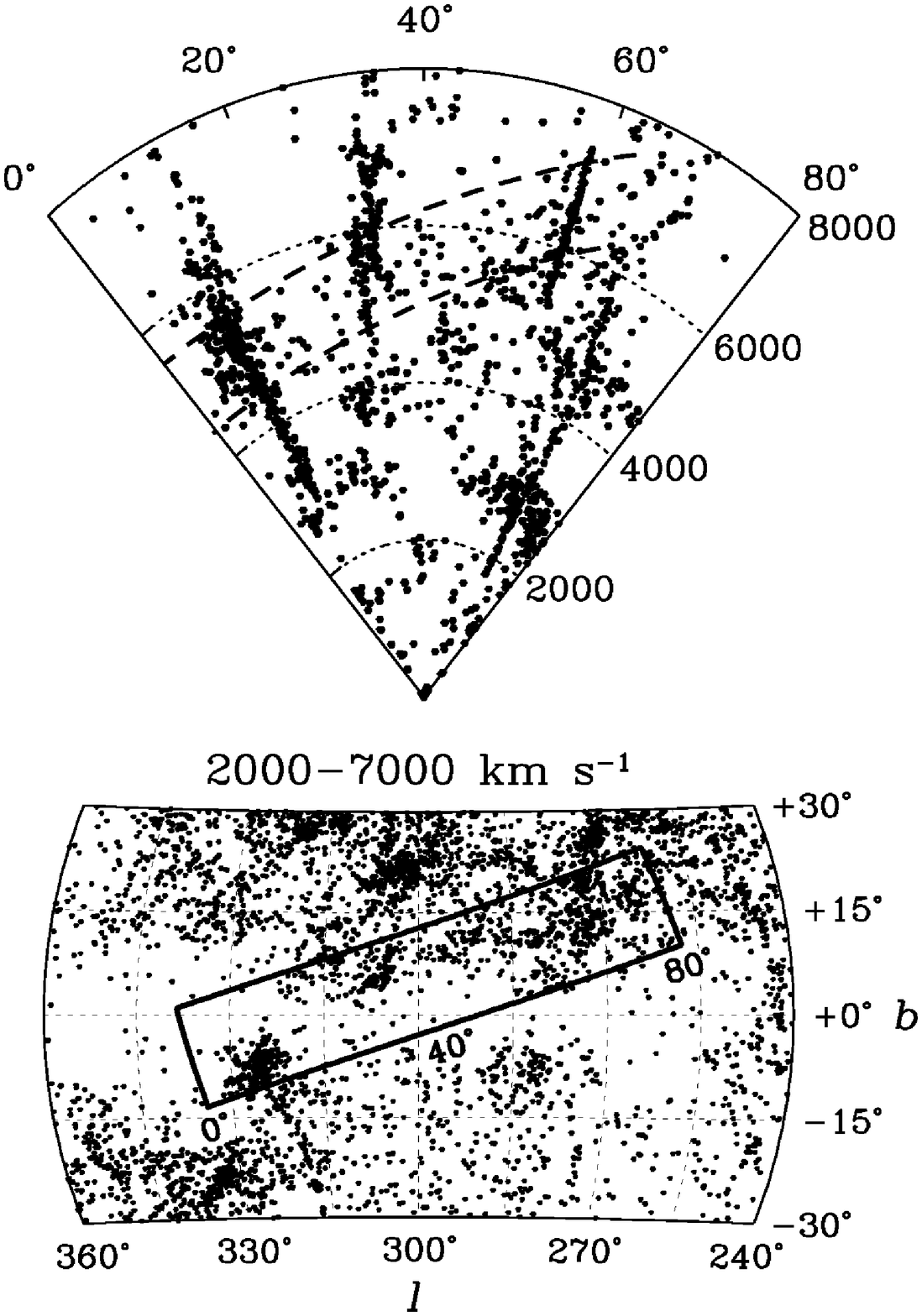}}
    \caption{The pieplot contains the galaxies in the
    80\deg\atimes15\deg~ rectangular strip shown in the Aitoff
    projection. The proposed Norma supercluster, seen as a trail of
    galaxies lying between the dashed lines, connects the
    `fingers-of-God' of the Norma cluster (11\deg, 4500\kms),
    CIZA~J1324.7--5736 (31\deg, 5570\kms) and Abell~S0639 (58\deg,
    6501\kms). The overdensity at (70\deg, 2800\kms) is the
    superposition of the Antlia cluster and the Hydra-Antlia extension
    seen in cross-section.}
    \label{fig:slice2}
  \end{center}
\end{figure}

Several large clusters are now known to reside in the GA region,
i.e. Norma, Pavo II, Centaurus, Hydra and CIZA~J1324.7--5736. However
the connections between these clusters are still poorly resolved. As
shown in Fig.~\ref{fig:sky}, the Pavo II and Norma clusters are
connected by a structure, which \cite{wou97} have suggested extends
through the ZoA towards the Cen-Crux overdensity. This connection is
highlighted by the noticeable peak around 5500\kms~in the combined
velocity distribution of non-cluster fields shown in the bottom panel
of Fig.~\ref{fig:histograms}. To examine this feature further,
Fig.~\ref{fig:slice1} and Fig.~\ref{fig:slice2} plot redshift slices
of the filament below and above the Galactic plane.

Evident in the foreground of the diagram in the upper panel of
Fig.~\ref{fig:slice1} is the Centaurus Wall. Appearing as a filament
of galaxies running across the sky at \cz~$\sim$~2600\kms, this
structure is separated by some 2000\kms~from the Norma structure. This
is in contradiction with earlier studies that have suggested the Norma
cluster is a nexus between the Centaurus Wall and the Norma
Supercluster \citep{wou97}. The dearth of galaxies in the ZoA is
clearly seen as the gap in the wall between the Norma and
CIZA~J1324.7--5736 clusters, which respectively appear as
`fingers-of-God' at 86\deg~and 108\deg~along the strip. However, below
the ZoA, the extent of the structure is clearly evident as the broad
wall of galaxies extends out from the Norma cluster, through the Pavo
II cluster and on towards higher redshifts. In the Aitoff projection
shown in the lower panel of Fig.~\ref{fig:slice1}, many additional,
smaller filaments are seen branching off from the main structure,
primarily at the location of the clusters. However a major branch
splits off at around $\sim$~(345\deg, --35\deg, 5000\kms) and
continues to $\sim$~(17\deg, \mbox{--22\deg}, 6000\kms). The main
filament appears to disperse at $\sim$~(5\deg, --50\deg, 5000\kms),
with apparent overdensities at greater galactic longitudes
(5\deg\lt\glong\lt30\deg, --60\deg\lt\glat\lt--45\deg) resulting from
the projection along the line of sight of clumps, including galaxies
in the Centaurus Wall.

Fig.~\ref{fig:slice2} shows a possible extension of the Norma
supercluster filament through the plane to higher galactic
latitudes. Here the progression to higher redshifts is hinted at as
the filament extends from the Norma Cluster (lying 11\deg~along the
strip), through CIZA~J1324.7--5736 (at 31\deg) and the Cen-Crux
feature (33\deg) and on towards Abell S0639 (58\deg). From this last
cluster an extension towards another overdensity located off the panel
at (268\deg, +17\deg, 9000\kms) may exist, but lack of redshifts makes
this difficult to discern. The Vela overdensity (280\deg, +6\deg,
6000\kms) lies next to Abell~S0639 and so forms a spur to the main
filament.  However, another intercluster connection from Abell S0639
appears to run at almost right angles to the Norma supercluster. This
filament extends through the overdensity located at (272\deg, +13\deg,
4500\kms), which is likely associated with Abell S0631 and Abell
S0628, both of which currently have no reported redshift, before
joining the Hydra cluster. As detailed in Section~\ref{sec:overview},
the large Hydra cluster is connected by the Hydra Wall to the
Centaurus cluster and by the Hydra-Antlia extension to the Antlia
cluster and galaxies at lower galactic latitudes.

Thus, from Abell S0639 to $\sim$(5\deg, --50\deg), there appears to
exist a continuous filament of galaxies stretching across
approximately 100\deg~(i.e. $\sim$\,120~Mpc) of the southern sky, with
a velocity dispersion $<400$\kms. From studies of inter-cluster
filaments in simulations, \cite*{col05} find a typical overdensity
along these structures of $\sim$~7 and cross-sectional radii of
$\sim$~2\,$h^{-1}$~Mpc. Thus, not including the associated clusters, a
filament of this size, dynamically centred at $\sim$~(325\deg,
--10\deg, 4800\kms), might contain a mass as high as
$\sim$~2.5\atimes10$^{15}$\msol. This is comparable to the mass of a
large cluster and so represents another potentially significant
component of the GA.

\section{Summary}
\label{sec:summary}

Using the 2dF on the AAT, we have measured 3053 redshifts in the
GA/SSC region, of which 2603 are new measurements. These redshifts
have helped reveal the composition of the GA, principally with the
resolution of the CIZA~J1324.7--5736/Cen-Crux feature. The X-ray
source is revealed to be a dynamically relaxed cluster with a mass
approximately 0.3--0.5 times that of the Norma Cluster, in
good agreement with previous estimates.

By combining the results of this survey with redshifts from the
literature, the major clusters associated with the GA are found to be
joined by a possibly wall-like structure. This filament extends from
Abell S0639, through the ZOA, where it meets the Norma cluster, and
continues down to $\sim$~(5\deg, --50\deg, 5000\kms). Together with
the Norma, Pavo II, CIZA~J1324.7--5736 and Abell S0639 clusters, we
can expect these structures to contribute a mass of
$\sim$10$^{16}$\msol~towards the GA.

We have also measured the masses and composition of several other
clusters behind the GA, including the Triangulum-Australis, Ara,
CIZA~J1514.6--4558 and CIZA~J1410.4--4246 clusters. These have been
proposed as possible sources to a continued flow beyond the GA. The
results from all these observations will be used in a subsequent paper
to model the flows in this complex and important region.

\section{Acknowledgements}

This paper is based on research taken with the 2dF at the AAT
telescope. We wish to thank all the AAO staff, in particular Scott
Croom and Rob Sharp, for their help and continued support of
2dF. DJR-S thanks PPARC for a research studentship. This research has
also made use of the NASA/IPAC Extragalactic Database (NED) which is
operated by the Jet Propulsion Laboratory, California Institute of
Technology, under contract with the National Aeronautics and Space
Administration and data products from 2MASS, which is a joint project
of the University of Massachusetts and the Infrared Processing and
Analysis Center/California Institute of Technology, funded by the
National Aeronautics and Space Administration and the National Science
Foundation.

\bibliographystyle{mn2e}
\bibliography{gastructures}

\begin{thebibliography}{}

\bibitem[\protect\citeauthoryear{{Aaronson} et~al.,}{{Aaronson}
  et~al.}{1989}]{aar89}
{Aaronson} M.,  et~al., 1989, ApJ, 338, 654

\bibitem[\protect\citeauthoryear{{Bardelli}, {Zucca}, {Zamorani}, {Moscardini}
  \& {Scaramella}}{{Bardelli} et~al.}{2000}]{bar00}
{Bardelli} S.,  {Zucca} E.,  {Zamorani} G.,  {Moscardini} L.,    {Scaramella}
  R.,  2000, MNRAS, 312, 540

\bibitem[\protect\citeauthoryear{{Barnes} et~al.,}{{Barnes}
  et~al.}{2001}]{bar01}
{Barnes} D.~G.,  et~al., 2001, MNRAS, 322, 486

\bibitem[\protect\citeauthoryear{{Branchini} et~al.,}{{Branchini}
  et~al.}{1999}]{bra99}
{Branchini} E.,  et~al., 1999, MNRAS, 308, 1

\bibitem[\protect\citeauthoryear{{Cappi} et~al.,}{{Cappi}
  et~al.}{1998}]{cap98}
{Cappi} A.,  et~al., 1998, A\&A, 336, 445

\bibitem[\protect\citeauthoryear{{Colberg}, {Krughoff} \& {Connolly}}{{Colberg}
  et~al.}{2005}]{col05}
{Colberg} J.~M.,  {Krughoff} K.~S.,    {Connolly} A.~J.,  2005, MNRAS, 359, 272

\bibitem[\protect\citeauthoryear{{Cole} et~al.,}{{Cole}  et~al.}{2001}]{col01b}
{Cole} S.,  et~al., 2001, MNRAS, 326, 255

\bibitem[\protect\citeauthoryear{{Colless} et~al.,}{{Colless}
  et~al.}{2001}]{col01}
{Colless} M.,  et~al., 2001, MNRAS, 328, 1039

\bibitem[\protect\citeauthoryear{{Cutri} et~al.,}{{Cutri}
  et~al.}{2003}]{cut03}
{Cutri} R.~M.,  et~al., 2003, VizieR Online Data Catalog, 2246

\bibitem[\protect\citeauthoryear{{Danese}, {de Zotti} \& {di Tullio}}{{Danese}
  et~al.}{1980}]{dan80}
{Danese} L.,  {de Zotti} G.,    {di Tullio} G.,  1980, A\&A, 82, 322

\bibitem[\protect\citeauthoryear{{Ebeling}, {Mullis} \& {Tully}}{{Ebeling}
  et~al.}{2002}]{ebe02}
{Ebeling} H.,  {Mullis} C.~R.,    {Tully} R.~B.,  2002, ApJ, 580, 774

\bibitem[\protect\citeauthoryear{{Ettori}, {Fabian} \& {White}}{{Ettori}
  et~al.}{1997}]{ett97}
{Ettori} S.,  {Fabian} A.~C.,    {White} D.~A.,  1997, MNRAS, 289, 787

\bibitem[\protect\citeauthoryear{{Faber} \& {Burstein}}{{Faber} \&
  {Burstein}}{1988}]{fab88}
{Faber} S.~M.,  {Burstein} D.,  1988, {Motions of galaxies in the neighborhood
  of the local group}.
Large-Scale Motions in the Universe: A Vatican study Week, pp 115--167

\bibitem[\protect\citeauthoryear{{Fairall}}{{Fairall}}{1998}]{fai98a}
{Fairall} A.~P.,  ed. 1998

\bibitem[\protect\citeauthoryear{{Fairall}, {Woudt} \&
  {Kraan-Korteweg}}{{Fairall} et~al.}{1998}]{fai98b}
{Fairall} A.~P.,  {Woudt} P.~A.,    {Kraan-Korteweg} R.~C.,  1998, A\&ASS, 127,
  463

\bibitem[\protect\citeauthoryear{{Hasegawa} et~al.,}{{Hasegawa}
  et~al.}{2000}]{has00}
{Hasegawa} T.,  et~al., 2000, MNRAS, 316, 326

\bibitem[\protect\citeauthoryear{{Heisler}, {Tremaine} \& {Bahcall}}{{Heisler}
  et~al.}{1985}]{hei85}
{Heisler} J.,  {Tremaine} S.,    {Bahcall} J.~N.,  1985, ApJ, 298, 8

\bibitem[\protect\citeauthoryear{{Henning}, {Kraan-Korteweg} \&
  {Stavely-Smith}}{{Henning} et~al.}{2005}]{hen05}
{Henning} P.~A.,  {Kraan-Korteweg} R.~C.,    {Stavely-Smith} L.,  2005, in ASP
  Conf. Ser. 329: Nearby Large-Scale Structures and the Zone of Avoidance
  p.~199

\bibitem[\protect\citeauthoryear{{H{\o}g}, {Fabricius}, {Makarov}, {Urban},
  {Corbin}, {Wycoff}, {Bastian}, {Schwekendiek} \& {Wicenec}}{{H{\o}g}
  et~al.}{2000}]{hog00}
{H{\o}g} E.,  {Fabricius} C.,  {Makarov} V.~V.,  {Urban} S.,  {Corbin} T.,
  {Wycoff} G.,  {Bastian} U.,  {Schwekendiek} P.,    {Wicenec} A.,  2000, A\&A,
  355, 27

\bibitem[\protect\citeauthoryear{{Huchra}, {Geller}, {Clemens}, {Tokarz} \&
  {Michel}}{{Huchra} et~al.}{1992}]{huc92}
{Huchra} J.~P.,  {Geller} M.~J.,  {Clemens} C.~M.,  {Tokarz} S.~P.,    {Michel}
  A.,  1992, Bull. CDS, 41, 31

\bibitem[\protect\citeauthoryear{{Hudson}}{{Hudson}}{1994}]{hud94}
{Hudson} M.~J.,  1994, MNRAS, 266, 468

\bibitem[\protect\citeauthoryear{{Hudson}, {Smith}, {Lucey} \&
  {Branchini}}{{Hudson} et~al.}{2004}]{hud04}
{Hudson} M.~J.,  {Smith} R.~J.,  {Lucey} J.~R.,    {Branchini} E.,  2004,
  MNRAS, 352, 61

\bibitem[\protect\citeauthoryear{{Jarrett}, {Chester}, {Cutri}, {Schneider},
  {Skrutskie} \& {Huchra}}{{Jarrett} et~al.}{2000}]{jar00}
{Jarrett} T.~H.,  {Chester} T.,  {Cutri} R.,  {Schneider} S.,  {Skrutskie} M.,
    {Huchra} J.~P.,  2000, AJ, 119, 2498

\bibitem[\protect\citeauthoryear{{Jones}, {Saunders}, {Read} \&
  {Colless}}{{Jones} et~al.}{2005}]{jon05}
{Jones} D.~H.,  {Saunders} W.,  {Read} M.,    {Colless} M.,  2005, PASA, 22,
  277

\bibitem[\protect\citeauthoryear{{Kaldare}, {Colless}, {Raychaudhury} \&
  {Peterson}}{{Kaldare} et~al.}{2003}]{kal03}
{Kaldare} R.,  {Colless} M.,  {Raychaudhury} S.,    {Peterson} B.~A.,  2003,
  MNRAS, 339, 652

\bibitem[\protect\citeauthoryear{{Kocevski}, {Ebeling}, {Mullis} \&
  {Tully}}{{Kocevski} et~al.}{2005}]{koc05}
{Kocevski} D.~D.,  {Ebeling} H.,  {Mullis} C.~R.,    {Tully} R.~B.,  2005,
  ArXiv:astro-ph/0512321

\bibitem[\protect\citeauthoryear{{Kocevski}, {Mullis} \& {Ebeling}}{{Kocevski}
  et~al.}{2004}]{koc04}
{Kocevski} D.~D.,  {Mullis} C.~R.,    {Ebeling} H.,  2004, ApJ, 608, 721

\bibitem[\protect\citeauthoryear{{Kolatt}, {Dekel} \& {Lahav}}{{Kolatt}
  et~al.}{1995}]{kol95}
{Kolatt} T.,  {Dekel} A.,    {Lahav} O.,  1995, MNRAS, 275, 797

\bibitem[\protect\citeauthoryear{{Koribalski}}{{Koribalski}}{2005}]{kor05}
{Koribalski} B.~S.,  2005, in ASP Conf. Ser. 329: Nearby Large-Scale Structures
  and the Zone of Avoidance p.~217

\bibitem[\protect\citeauthoryear{{Koribalski} et~al.,}{{Koribalski}
  et~al.}{2004}]{kor04}
{Koribalski} B.~S.,  et~al., 2004, AJ, 128, 16

\bibitem[\protect\citeauthoryear{{Kraan-Korteweg}, {Fairall} \&
  {Balkowski}}{{Kraan-Korteweg} et~al.}{1995}]{kra95}
{Kraan-Korteweg} R.~C.,  {Fairall} A.~P.,    {Balkowski} C.,  1995, A\&A, 297,
  617

\bibitem[\protect\citeauthoryear{{Kraan-Korteweg}, {Ochoa}, {Woudt} \&
  {Andernach}}{{Kraan-Korteweg} et~al.}{005a}]{kra05a}
{Kraan-Korteweg} R.~C.,  {Ochoa} M.,  {Woudt} P.~A.,    {Andernach} H.,  2005a,
  in {Fairall} A.~P.,  {Woudt} P.~A.,  eds, {ASP Conf. Ser. Vol. 329, Nearby
  Large-Scale Structures and the Zone of Avoidance. Astron. Soc. Pac., San
  Francisco,} pp 159--165

\bibitem[\protect\citeauthoryear{{Kraan-Korteweg}, {Staveley-Smith}, {Donley},
  {Koribalski} \& {Henning}}{{Kraan-Korteweg} et~al.}{005b}]{kra05b}
{Kraan-Korteweg} R.~C.,  {Staveley-Smith} L.,  {Donley} J.,  {Koribalski} B.,
   {Henning} P.~A.,  2005b, in {Colless} M.,  {Staveley} S.,   {Stathakis} R.,
  eds, ASP Conf. Ser. Vol. 216, Maps of the Cosmos pp 203--210

\bibitem[\protect\citeauthoryear{{Kraan-Korteweg} \& {Woudt}}{{Kraan-Korteweg}
  \& {Woudt}}{1994}]{kra94}
{Kraan-Korteweg} R.~C.,  {Woudt} P.~A.,  1994, in ASP Conf. Ser. 67: Unveiling
  Large-Scale Structures Behind the Milky Way p.~89

\bibitem[\protect\citeauthoryear{{Kraan-Korteweg} \& {Woudt}}{{Kraan-Korteweg}
  \& {Woudt}}{1999}]{kra99}
{Kraan-Korteweg} R.~C.,  {Woudt} P.~A.,  1999, PASA, 16, 53

\bibitem[\protect\citeauthoryear{{Kraan-Korteweg}, {Woudt}, {Cayatte},
  {Fairall}, {Balkowski} \& {Henning}}{{Kraan-Korteweg} et~al.}{1996}]{kra96}
{Kraan-Korteweg} R.~C.,  {Woudt} P.~A.,  {Cayatte} V.,  {Fairall} A.~P.,
  {Balkowski} C.,    {Henning} P.~A.,  1996, Nat, 379, 519

\bibitem[\protect\citeauthoryear{{Lahav}, {Yamada}, {Scharf} \&
  {Kraan-Korteweg}}{{Lahav} et~al.}{1993}]{lah93}
{Lahav} O.,  {Yamada} T.,  {Scharf} C.,    {Kraan-Korteweg} R.~C.,  1993,
  MNRAS, 262, 711

\bibitem[\protect\citeauthoryear{{Laustsen}, {Schuster} \& {West}}{{Laustsen}
  et~al.}{1977}]{lau77}
{Laustsen} S.,  {Schuster} H.-E.,    {West} R.~M.,  1977, A\&A, 59, 3

\bibitem[\protect\citeauthoryear{{Lucey}, {Radburn-Smith} \& {Hudson}}{{Lucey}
  et~al.}{2005}]{luc05}
{Lucey} J.,  {Radburn-Smith} D.,    {Hudson} M.,  2005, in ASP Conf. Ser. 329:
  Nearby Large-Scale Structures and the Zone of Avoidance p.~21

\bibitem[\protect\citeauthoryear{{Lucey} \& {Carter}}{{Lucey} \&
  {Carter}}{1988}]{luc88}
{Lucey} J.~R.,  {Carter} D.,  1988, MNRAS, 235, 1177

\bibitem[\protect\citeauthoryear{{Lynden-Bell}, {Faber}, {Burstein}, {Davies},
  {Dressler}, {Terlevich} \& {Wegner}}{{Lynden-Bell} et~al.}{1988}]{lyn88}
{Lynden-Bell} D.,  {Faber} S.~M.,  {Burstein} D.,  {Davies} R.~L.,  {Dressler}
  A.,  {Terlevich} R.~J.,    {Wegner} G.,  1988, ApJ, 326, 19

\bibitem[\protect\citeauthoryear{{Mathewson}, {Ford} \& {Buchhorn}}{{Mathewson}
  et~al.}{1992}]{mat92}
{Mathewson} D.~S.,  {Ford} V.~L.,    {Buchhorn} M.,  1992, ApJL, 389, 5

\bibitem[\protect\citeauthoryear{{Meyer} et~al.,}{{Meyer}
  et~al.}{2004}]{mey04}
{Meyer} M.~J.,  et~al., 2004, MNRAS, 350, 1195

\bibitem[\protect\citeauthoryear{{Mills}}{{Mills}}{1952}]{mil52}
{Mills} B.~Y.,  1952, Australian J.~Sci.~Res.~serie A 5, 5, 266-287 (1952), 5,
  266

\bibitem[\protect\citeauthoryear{{Mullis}, {Ebeling}, {Kocevski} \&
  {Tully}}{{Mullis} et~al.}{2005}]{mul05}
{Mullis} C.~R.,  {Ebeling} H.,  {Kocevski} D.~D.,    {Tully} R.~B.,  2005, in
  ASP Conf. Ser. 329: Nearby Large-Scale Structures and the Zone of Avoidance
  p.~183

\bibitem[\protect\citeauthoryear{{Nagayama} et~al.,}{{Nagayama}
  et~al.}{2004}]{nag04}
{Nagayama} T.,  et~al., 2004, MNRAS, 354, 980

\bibitem[\protect\citeauthoryear{{Nagayama}, {Nagata}, {Sato}, {Woudt} \&
  {Irsf/Sirius Team}}{{Nagayama} et~al.}{2005}]{nag05}
{Nagayama} T.,  {Nagata} T.,  {Sato} S.,  {Woudt} P.~A.,    {Irsf/Sirius Team}
  2005, in ASP Conf. Ser. 329: Nearby Large-Scale Structures and the Zone of
  Avoidance p.~177

\bibitem[\protect\citeauthoryear{{Proust} et~al.,}{{Proust}
  et~al.}{2005}]{pro05}
{Proust} D.,  et~al., 2005, arXiv:astro-ph/0509903

\bibitem[\protect\citeauthoryear{{Raychaudhury}}{{Raychaudhury}}{1989}]{ray89}
{Raychaudhury} S.,  1989, Nat, 342, 251

\bibitem[\protect\citeauthoryear{{Reisenegger}, {Quintana}, {Carrasco} \&
  {Maze}}{{Reisenegger} et~al.}{2000}]{rei00}
{Reisenegger} A.,  {Quintana} H.,  {Carrasco} E.~R.,    {Maze} J.,  2000, AJ,
  120, 523

\bibitem[\protect\citeauthoryear{{Rines}, {Geller}, {Kurtz} \&
  {Diaferio}}{{Rines} et~al.}{2003}]{rin03}
{Rines} K.,  {Geller} M.~J.,  {Kurtz} M.~J.,    {Diaferio} A.,  2003, AJ, 126,
  2152

\bibitem[\protect\citeauthoryear{{Rowan-Robinson} et~al.,}{{Rowan-Robinson}
  et~al.}{2000}]{row00}
{Rowan-Robinson} M.,  et~al., 2000, MNRAS, 314, 375

\bibitem[\protect\citeauthoryear{{Royston}}{{Royston}}{1995}]{roy95}
{Royston} P.,  1995, Applied Statistics, 44, 547

\bibitem[\protect\citeauthoryear{{Scaramella}, {Baiesi-Pillastrini},
  {Chincarini}, {Vettolani} \& {Zamorani}}{{Scaramella} et~al.}{1989}]{sca89}
{Scaramella} R.,  {Baiesi-Pillastrini} G.,  {Chincarini} G.,  {Vettolani} G.,
   {Zamorani} G.,  1989, Nature, 338, 562

\bibitem[\protect\citeauthoryear{{Schr\"oder}, {Kraan-Korteweg}, {Mamon} \&
  {Woudt}}{{Schr\"oder} et~al.}{2005}]{sch05}
{Schr\"oder} A.~C.,  {Kraan-Korteweg} R.~C.,  {Mamon} G.~A.,    {Woudt} P.~A.,
  2005, in {Fairall} A.~P.,  {Woudt} P.~A.,  eds, {ASP Conf. Ser. Vol. 329,
  Nearby Large-Scale Structures and the Zone of Avoidance. Astron. Soc. Pac.,
  San Francisco,} pp 167--176

\bibitem[\protect\citeauthoryear{{Shapiro} \& {Wilk}}{{Shapiro} \&
  {Wilk}}{1965}]{sha65}
{Shapiro} S.~S.,  {Wilk} M.~B.,  1965, Biometrika, 52, 591

\bibitem[\protect\citeauthoryear{{Smith}, {Hudson}, {Lucey}, {Schlegel} \&
  {Davies}}{{Smith} et~al.}{2000}]{smi00}
{Smith} R.~J.,  {Hudson} M.~J.,  {Lucey} J.~R.,  {Schlegel} D.~J.,    {Davies}
  R.~L.,  2000, in Astronomical Society of the Pacific Conference Series p.~39

\bibitem[\protect\citeauthoryear{{Stein}}{{Stein}}{1994}]{ste94}
{Stein} P.,  1994, Ph.D.~thesis, University of Basel

\bibitem[\protect\citeauthoryear{{Stein}}{{Stein}}{1996}]{ste96}
{Stein} P.,  1996, A\&ASS, 116, 203

\bibitem[\protect\citeauthoryear{{Stein}}{{Stein}}{1997}]{ste97}
{Stein} P.,  1997, A\&A, 317, 670

\bibitem[\protect\citeauthoryear{{Tamura}, {Fukazawa}, {Kaneda}, {Makishima},
  {Tashiro}, {Tanaka} \& {Bohringer}}{{Tamura} et~al.}{1998}]{tam98}
{Tamura} T.,  {Fukazawa} Y.,  {Kaneda} H.,  {Makishima} K.,  {Tashiro} M.,
  {Tanaka} Y.,    {Bohringer} H.,  1998, PASJ, 50, 195

\bibitem[\protect\citeauthoryear{{Tashiro} et~al.,}{{Tashiro}
  et~al.}{1998}]{tas98}
{Tashiro} M.,  et~al., 1998, ApJ, 499, 713

\bibitem[\protect\citeauthoryear{{Tonry} \& {Davis}}{{Tonry} \&
  {Davis}}{1979}]{ton79}
{Tonry} J.,  {Davis} M.,  1979, AJ, 84, 1511

\bibitem[\protect\citeauthoryear{{Tonry}, {Blakeslee}, {Ajhar} \&
  {Dressler}}{{Tonry} et~al.}{2000}]{ton00}
{Tonry} J.~L.,  {Blakeslee} J.~P.,  {Ajhar} E.~A.,    {Dressler} A.,  2000,
  ApJ, 530, 625

\bibitem[\protect\citeauthoryear{{Wakamatsu}, {Malkan}, {Nishida}, {Parker},
  {Saunders} \& {Watson}}{{Wakamatsu} et~al.}{2005}]{wak05}
{Wakamatsu} K.,  {Malkan} M.~A.,  {Nishida} M.~T.,  {Parker} Q.~A.,  {Saunders}
  W.,    {Watson} F.~G.,  2005, in {Fairall} A.~P.,  {Woudt} P.~A.,  eds, {ASP
  Conf. Ser. Vol. 329, Nearby Large-Scale Structures and the Zone of Avoidance.
  Astron. Soc. Pac., San Francisco,} pp 189--198

\bibitem[\protect\citeauthoryear{{West} \& {Tarenghi}}{{West} \&
  {Tarenghi}}{1989}]{wes89}
{West} R.~M.,  {Tarenghi} M.,  1989, A\&A, 223, 61

\bibitem[\protect\citeauthoryear{{Woudt}}{{Woudt}}{1998}]{wou98}
{Woudt} P.~A.,  1998, Ph.D.~thesis, University of Cape Town

\bibitem[\protect\citeauthoryear{{Woudt}, {Fairall} \&
  {Kraan-Korteweg}}{{Woudt} et~al.}{1997}]{wou97}
{Woudt} P.~A.,  {Fairall} A.~P.,    {Kraan-Korteweg} R.~C.,  1997, in ASP Conf.
  Ser. Vol. 117, Dark and Visible Matter in Galaxies and Cosmological
  Implications pp 373--379

\bibitem[\protect\citeauthoryear{{Woudt}, {Kraan-Korteweg}, {Cayatte},
  {Balkowski} \& {Felenbok}}{{Woudt} et~al.}{2004}]{wou04}
{Woudt} P.~A.,  {Kraan-Korteweg} R.~C.,  {Cayatte} V.,  {Balkowski} C.,
  {Felenbok} P.,  2004, A\&A, 415, 9

\bibitem[\protect\citeauthoryear{{Woudt}, {Kraan-Korteweg} \&
  {Fairall}}{{Woudt} et~al.}{1999}]{wou99}
{Woudt} P.~A.,  {Kraan-Korteweg} R.~C.,    {Fairall} A.~P.,  1999, A\&A, 352,
  39

\end{thebibliography}

\label{lastpage}
\end{document}